# Training Evaluation in a Smart Farm using Kirkpatrick Model: A Case Study of Chiang Mai

Suepphong Chernbumroong
College of Arts, Media and Techology,
Chiang Mai University,
Chiang Mai, Thailand
suepphong@camt.info

Pradorn Sureephong
College of Arts, Media and Techology,
Chiang Mai University,
Chiang Mai, Thailand
dorn@camt.info

Paweena Suebsombut
School of Information Technology,
Mae Fah Luang University,
Chiang Rai, Thailand
paweena.sue@mfu.ac.th

Aicha Sekhari
Decision and Information Sciences for Production Systems,
University Lumiere Lyon 2,
Lyon, France
aicha.sekhari@univ-lyon2.fr

*Abstract*— Farmers can now use IoT to improve farm efficiency and productivity by using sensors for farm monitoring to enhance decision-making in areas such as fertilization, irrigation, climate forecast, and harvesting information. Local farmers in Chiang Mai, Thailand, on the other hand, continue to lack knowledge and experience with smart farm technology. As a result, the *'SUNSpACe'* project, funded by the European Union's Erasmus+ Program, was launched to launch a training course which improve the knowledge and performance of Thai farmers. To assess the effectiveness of the training, The Kirkpatrick model was used in this study. Eight local farmers took part in the training, which was divided into two sections: mobile learning and smart farm laboratory. During the training activities, different levels of the Kirkpatrick model were conducted and tested: reaction (satisfaction test), learning (knowledge test), and behavior (performance test). The overall result demonstrated the participants' positive reaction to the outcome. The paper also discusses the limitations and suggestions for training activities.

*Keywords*— *Smart Farmers, Kirkpatrick Model, Training Evaluation*

## I. Introduction (Heading 1)

Farming is essential sources of food for the world's population. Farming depends on the environment and weather conditions. During this period, the world's agriculture has been greatly affected by the volatile weather conditions, for example, climate change, which has an effect on farming and food production. For sustainability of food production, it is necessary to adopt smart technologies for more precision farming and make decisions in time for maximum resource utilization [1].

Current smart technologies consist of artificial intelligence (AI), internet of things (IoT), and robotics with transformation of data, automatic systems and control, for example, smart health care, smart home, etc., and now it is initialed to be adopted in the farming sector called "Smart Farming," which is the combination of ICT (information and communication) technologies into equipment and sensors for food production and crop cultivation processes. Smart Farming technologies include IoT, robotics, AI, tracking systems using GPS, etc. They enable farmers to monitor and allocate data to particular farm locations, automating and controlling systems. Robotics and AI enable highly precise monitoring and control systems. A traking systemwith GPS helps to track animals or enhance the self-driving of agricultural machines more efficiently and precisely. IoT and sensors measure farm and crop environment such as soil moisture, air temperature, humidity, and so on to support efficient decision-making, including crop disease infestation sensors [2]. Furthermore, anticipatory planning and process optimization are the most important aspects of smart farming in terms of efficiency and precision [3].

Nowadays, some farmers can use IoT technology with sensors to monitor their farm for their decision-making improvement and enhance their farm practice and management efficiently, such as weather forecasts, fertilization, irrigation, etc. [4]. This furtherance enables farmers to make better farm management decisions by utilizing available resources more efficiently, resulting in appropriate yields and higher income [5].

However, local farmers in Chiang Mai province, Thailand, still lack knowledge of smart farm technology. As a result, smart farm technology training courses are very important for transferring smart farm technology knowledge to farmers, which is one goal of the SUstainable developmeNt Smart Agriculture Capacity (SUNSpACe) project, which proposes to create an online learning platform for transferring smart farm technology knowledge to farmers. The SUNSpACe project, which stands for Sustainable Development Smart Agriculture Capacity, provides an education and training system to assist farmers in understanding the use and utility of new technologies. The overall goal of developing a smart farm knowledge management system in Asian countries is to support socioeconomic growth in three target countries: Thailand, Nepal, and Bhutan.

Following the completion of the training course, it will be implemented in the Asian partner countries (Thailand, Nepal, and Bhutan). In this paper, we use the Kirkpatrick model to assess the proposed training course. As a result, the goal of this paper is to propose a Kirkpatrick model to evaluate the training course and to reveal the implementation results of the Kirkpatrick model in order to examine the SUNSpACe project's proposed Kirkpatrick model performance.

## II. SUNSpACe Kirkpatrick Model

Earlier, several studies were conducted in conjunction with the SUNSpACe project from 2019 to 2021. Consider the smart farming literacy study, which sought to comprehend farming activity, as well as farmers' experience and ability with smart farming, in order to design an appropriate learning tools for farmers to improve farming production [6]. The findings confirmed that Asian farmers are interested in technological devices. Meanwhile, a survey of smart agriculture literacy in aspect of farmers' experience and skills specific to Chiang Mai and Khon Kaen provinces in Thailand [7] revealed that farmers in Chiang Mai and Khon Kaen provinces have vastly different levels of smart agriculture literacy. As a result, the training section should take the participants' backgrounds into account. Because the training section involves mobile learning, the learning content on the online platform must be carefully designed. Users' reactions to MOOCs vary, according to [8]. The low completion rate in MOOCs is a major source of concern. The user's learning style and technology usage must be investigated for a higher completion rate and learning outcome. As previously stated, most previous studies attempted to understand only the users' backgrounds and preferences. However, we discovered that the perspectives on learning performance and training outcomes were still missing. As a result, Kirkpatrick was used in this study to assess the effectiveness of training activities.

Kirkpatrick's approach to evaluation was first proposed in 1959. As part of its semi–centennial celebrations [9,10], the model was extensively reviewed. It is made up of four evaluation levels that's are used to assess workplace training. The first level is "reaction", which assesses how participants in a training program respond to it. The second level is learning to assess how far students "skills, knowledge, or attitudes have advanced. The third level is learner behavior that has changed as a result of the training program. At this level, evaluate attempts to answer this question: Are the newly acquired skills, knowledge, or attitudes being used in the learner's everyday environment? The fourth level is results, which assess the training program's success [11]. The Kirkpatrick model's levels of reaction, learning, and behavior were used as an evaluation method for our training course in the SUNSpACe project (see table I). The only result level that is not used in this phase is due to our projects conditions and constraints.

TABLE I. PROPOSED SUNSPACE KIRKPATRICK MODEL

| Training Section | | Mobile Learning | Smart Farm Laboratory |
|---|---|---|---|
| Kirkpatrick | Reaction Level | Satisfaction Test | Satisfaction Test |
| | Learning Level | Knowledge Test | - |
| | Behavior Level | - | Performance Test |
| | Result Level | - | - |
| Bloom Taxonomy | Knowledge dimension | Remembering and Understanding | Apply |

The various training sections have been designed and developed in accordance with the course training. The first section is about mobile learning, which is based on the concept of MOOCs (Massive open online courses). The fundamentals of smart farming are presented in the form of learning videos available online. According to Bloom's taxonomy (knowledge dimension) [9], the project hoped that mobile learning would encourage farmers to gain general knowledge of smart farming in terms of remembering and understanding skills. Thus, in this section, the levels of reaction (satisfaction test) and learning (knowledge test) are used. Meanwhile, the second section is a smart farm laboratory where farmers were taught how to use IoT equipment and interpret basic data. This section will allow the farmer to hone their application skills. As a result, the training outcome is evaluated using reaction level (satisfaction) and behavior (performance test).

### III. METHODOLOGY

#### A. Learning Content

This study's learning material was divided into two sections: the SUNSpACe MOOC and the Smart farm Laboratory. The SUNSpACe MOOC's content is video based, with instructor-guided and PowerPoint lessons covering a variety of topics. The presentations in the videos included images, diagrams, and the lecturer's voiceover via PowerPoint, and each video clip lasted 5-8 minutes (see table II). This curriculum has a total run time of 120 minutes. At the end of each video clip, a multiple-choice exercise was given. All of the quizzes were required of all participants.

TABLE II. LEARNING VIDEO IN MOBILE LEARNING

| Topic | Subtopic | Duration |
|---|---|---|
| Sensor Introduction | Water sensor, Soil sensor, Weather station, Air sensor | 10 minutes |
| Sensor Installation | Water sensor, Soil sensor, Weather station, Air sensor | 30 minutes |
| Sensor Adoption | Water sensor, Soil sensor, Weather station, Air sensor | 20 minutes |
| Data Interpretation | Soil moisture and temperature, Water temperature, Air temperature and Air relative humidity, Weather | 30 minutes |
| Irrigation Control | Controller | 30 minutes |

SUNSpACe MOOC is an online platform developed by the project with the goal of allowing an open access and unlimited participation through a mobile application. This platform offers an interactive course with user forums to help farmers, researchers, the instructor, and field experts build learning communities. The platform also supports basic digital content such as audio, videos, images, XML files, and online documents, as well as instant feedback on assignments such as multiple-choice questions or quick quizzes. This application is also available on Android and iOS.

In the SUNSpACe project, the smart farm laboratory serves as a demonstration site where farmers can learn about various types of IoT sensors equipment used in farming, such as soil sensors, water sensors, air sensors, and weather stations. Furthermore, the laboratory allows for the stimulation and training of all SUNSpACe farm equipment. (Both hardware and software). Participants were given the opportunity to visit various learning stations on the project farm.

## B. Participant and Procedure

Eight local farmers were invited to participate in the experiment. All participants volunteered and were informed that they would be taking part in the SUNSpACe workshop. The participants' average age is 44 years old (male = 6, female = 2). They all have bachelor's degrees, and the average farm experience is four years. Before being assigned to enroll in and complete all of the learning content of the SUNSpACe MOOC, all participants were asked to complete the pre-test (20 minutes). Participants must have completed the curriculum by watching all of the video clips and passing al of the quizzes in order to complete it. Participants had one day to complete the curriculum. They could learn on their own mobile devices, at their home or at workplace, and at their own pace. There were no repercussions for participants who did not complete the curriculum. Following completion of all tasks, participants were asked to complete a post-test (20 minute) satisfaction questionnaire on the same platform. All participants were invited to a performance test in the SUNSpACe smart farm laboratory seven days later.

## C. Evaluation

The evaluation method used in this study was based on the Kirkpatrick model's four levels. The satisfaction questionnaire was developed as a tool for measuring the level of reaction. The questionnaire was divided into two sections: a study on the MOOC platform and learning content (12 items) and a smart farm laboratory (6 items), with a 5-point Likert scale ranging from 1 (strongly disagree) to 5 (strongly agree) [10]. The second level is learning, which aims to assess participants' knowledge and skills. As a result, we use the pre- and post-test as a tool for knowledge testing. Multiple-choice assessment sheet (A, B, C, and D). All questions were to be completed by the participants. There are a total of 18 questions (11 for remembering and 7 for understanding). Remembering is recalling a fact or a basic concept, whereas understanding is explaining an idea or a concept: "What is the definition of relative humidity?" is an example of a remembering question. While an example of an understanding question is "Which of the following sentences is not an activity that raises the temperature of the water?" The final level is measured by the performance test (unable to do = 0, fair = 1, do well = 2).

## IV. RESULTS

Based on the training section, the outcome was divided into two parts. The first component is mobile learning, which began with all volunteer participants enrolling in MOOCs at the start of the study. The "SUNSpACe MOOC" mobile application ran without bugs and smoothly, and no issues were reported by participants. They could access all of the app's features, including all of the videos and quizzes. Participants had one day to complete all of the learning content in the MOOC, as stated in the methodology section. We discovered that the majority of the participants had a positive reaction at the end of the session.

According to the results of the learning level (see Figure 1), all participants' knowledge has improved (pre- test average score = 7.25, post-test average score = 13). Participants made positive progress based on their ability to remember and comprehend test results. The average pre-test score for remembering is 3.75, and the post-test score is 7.12 (progress = 3.37), whereas the average pre- test score for understanding is 3.5, and the post-test score is 5.87 (progress = 2.37).

However, we discovered that only two of the participants with the lowest scores had issues with mobile learning assessment.

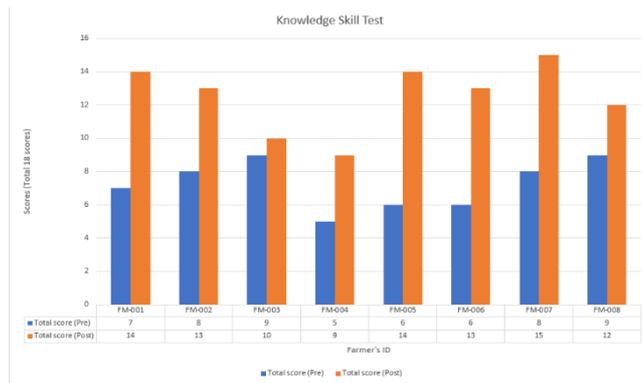

Fig. 1. The result of knowledge test in mobile learning section

TABLE III. THE RESULT OF SATISFACTION TEST IN MOBILE LEARNING SECTION

| Item | Score | Level |
| --- | --- | --- |
| S1.1 | 4.5 | Strongly Agree |
| S1.2 | 3.87 | Agree |
| S1.3 | 3.37 | Neutral |
| S1.4 | 4.87 | Strongly Agree |
| S1.5 | 4.5 | Strongly Agree |
| S1.6 | 4.75 | Strongly Agree |
| S1.7 | 4.62 | Strongly Agree |
| S1.8 | 4.75 | Strongly Agree |
| S1.9 | 3.62 | Agree |
| S1.10 | 3.25 | Neutral |
| S1.11 | 4.62 | Strongly Agree |
| S1.12 | 4.62 | Strongly Agree |

The result from satisfaction test (see table III) in mobile learning section demonstrated that participants were satisfied with the learning activity (the average score satisfaction is 4.28). However, only the items concerned with the learning content from a mobile phone perspective (S1.3) and device installation difficulty (S1.10) meet the neutral level. According to the results of the knowledge and satisfaction tests, we discovered that some participants between the ages of 55 and 60 had difficulty with mobile device training. As a result, we propose that the user interface (UI) of the software design in MOOC platforms (text size, color, symbol) should consider mobile responsiveness for users of all ages.

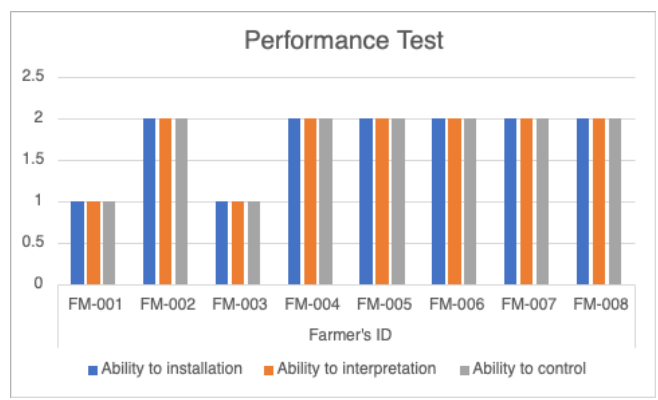

Fig. 2. The result of performance test in smart farm laboratory

The second component is the Smart farm Laboratory, where participants were trained and evaluated (see Figure 2). Overall, the performance test indicated that six participants (75 percent), performed "well" in all tasks: installation, interpretation, and control. The training course clearly aids participants in acquiring new skills and knowledge that influence their behavior. However, only two people (25%) scored "fair.

TABLE IV. THE RESULT OF SATISFACTION TEST IN SMART FARM LABORATORY

| Item | Score | Level |
|---|---|---|
| S2.1 | 4.65 | Strongly Agree |
| S2.2 | 4.65 | Strongly Agree |
| S2.3 | 4.75 | Strongly Agree |
| S2.4 | 4.62 | Strongly Agree |
| S2.5 | 3.75 | Agree |
| S2.6 | 3.35 | Neutral |

Participants who enrolled in the smart farm laboratory were extremely satisfied with the training course (the average satisfaction score is 4.29). (See table IV). Only item S2.6 received a neutral level of satisfaction because participants were concerned the training was too short. Based on the results of the performance tests and participant satisfaction in smart laboratory training, we believed that extending the future duration of smart farm training from 1-4 weeks would improve participant performance and satisfaction because the training activities are conducted both online and on-site.

V. CONCLUSION

The goal of this study is to use the Kirkpatrick model to assess the outcome of a training course in the SUNSpACe project. Eight local farmers were invited to participate in the training, which was divided into two sections (Mobile learning and smart farm laboratory). The results of the reaction and the learning levels in the mobile learning section confirmed that the SUNSpACe MOOC helps all participants improve their knowledge. However, some older participants are still experiencing issues with the platforms current graphic design, which requires revision, even though the training duration is limited, the participants performance in applying skills improved, resulting in a smart farm laboratory section. As a result, the project intends to improve and extend the duration of the training activities in the near future.

ACKNOWLEDGMENT

The authors would like to express gratitude to the Knowledge and Innovation Research Laboratory (KIRLY), College of Arts, Media and Technology, Chiang Mai University, Thailand which provided invaluable assistance throughout the study. In addition, we would like to thank al the local farmers who volunteered to take part in the study. The SUstainable developmeNt Smart Agriculture Capacity (SUNSpACe) project is co-funded by the Erasmus+ Program of the European Union, reference 598748-EPP-1-2018-1-FR-EPPKA2-CBHE-JP (2018-3228/001-001).

REFERENCES

[1] Wheeler T, von Braun J (2013) Climate change impacts on global food security. Science 341 (80):508–513.
https://doi.org/10.1126/science.1239402

[2] Weltzien/Gebbers, 2016, Aktueller Stand der Technik im Bereich der Sensoren für Precision Agriculture in: Ruckelshausen et al., Intelligente Systeme Stand der Technik und neue Möglichkeiten, pp. 16 et seqq

[3] Schönfeld, Max V., Reinhard Heil, and Laura Bittner. "Big data on a farm—Smart farming." Big Data in Context (2018): 109-120.

[4] Pivoto D, Waquil PD, Talamini E, Finocchio CPS, Dalla Corte VF, de Vargas Mores G (2018) Scientific development of smart farming technologies and their application in Brazil. Inf Process Agric 5:21–32. https://doi.org/10.1016/J.INPA.2017.12.002

[5] Supreetha MA, Mundada MR, Pooja JN (2019) Design of a smart water-saving irrigation system for agriculture based on a wireless sensor network for better crop yield. 93–104.
https://doi.org/10.1007/978-981-13-0212-1_11

[6] U. Yangchen et al., "Preliminary study on Smart farming literacy: A case study in Barp gewog, Punakha District, Bhutan," 2021 Joint International Conference on Digital Arts, Media and Technology with ECTI Northern Section Conference on Electrical, Electronics, Computer and Telecommunication Engineering, 2021, pp. 340-345, doi: 10.1109/ECTIDAMTNCON51128.2021.9425772.

[7] P. Suebsombut, S. Chernbumroong, P. Sureephong, P. Jaroenwanit, P. Phuensane and A. Sekhari, "Comparison of Smart Agriculture Literacy of Farmers in Thailand," 2020 Joint International Conference on Digital Arts, Media and Technology with ECTI Northern Section Conference on Electrical, Electronics, Computer and Telecommunications Engineering (ECTI DAMT & NCON), 2020, pp. 242-245, doi: 10.1109/ECTIDAMTNCON48261.2020.9090695.

[8] Chernbumroong, Suepphong, Pradorn Suereephong, and Kitti Puritat. "Massive Open Online Course Related Learning Style and Technology Usage Patterns of Thai Tourism Professionals." International Journal of Emerging Technologies in Learning 12.11 (2017).

[9] Kirkpatrick, J. & Kayser-Kirkpatrick,W. (2014). The Kirkpatrick four levels: A fresh look after 55 years. Ocean City: Kirkpatrick Partners.

[10] Paull, Megan, Craig Whitsed, and Antonia Girardi. "Applying the Kirkpatrick model: Evaluating an'interaction for learning framework'curriculum intervention." Issues in Educational Research 26, no. 3 (2016): 490.

[11] Kirkpatrick, James D., and Wendy Kayser Kirkpatrick. Kirkpatrick's four levels of training evaluation. Association for Talent Development, 2016.

[12] Anderson, L. W., & Krathwohl, D. R. (2001). A taxonomy for learning, teaching, and assessing: A revision of Bloom's taxonomy of educational objectives. New York: Longman.

[13] Likert, Rensis (1932). "A Technique for the Measurement of Attitudes". Archives of Psy-chology. 140: 1–5